\documentclass[12pt]{article}

\usepackage{graphicx}
\usepackage{amsfonts}
\usepackage[top=2.75cm, bottom=2.75cm, left=2.75cm, right=2.75cm]{geometry} 

\def\+{{+\!\!\!+}} 
\def\pp{\mbox{\tiny${}_{\stackrel\+ =}$}}

\def\d{\partial}

\def\G{\Gamma}

\def\p{\psi} 
 
\def\e{\varepsilon}

\def\th{\theta}

\def\pmb#1{\setbox0=\hbox{#1}% 
\kern.0em\copy0\kern-\wd0 
\kern-.04em\copy0\kern-\wd0 
\kern.08em\copy0\kern-\wd0 
\kern-.04em\raise.0433em\box0 }         %poor man's bold macro (TexBook) 
\def\half{\textstyle{1\over 2}}

\newcommand{\nc}{\newcommand} 
\nc{\beq}{\begin{equation}} 
\nc{\eeq}[1]{\label{#1}\end{equation}} 
\nc{\ber}{\begin{eqnarray}} 
\nc{\eer}[1]{\label{#1}\end{eqnarray}} 
\nc{\pek}[1]{\cite{#1}} 
\nc{\enr}[1]{(\ref{#1})} 
\nc{\kal}[1]{{\cal{#1}}} 
\nc{\dott}{\;\cdot\;} 
\nc{\coker}{\mathrm{coker}}
\nc{\ie}{{\it i.e.}}
\nc{\eg}{{\it e.g.}}
\def\ph#1{\phantom{#1}}
\def\p{{(+)}}
\def\m{{(-)}}
\def\ppm{{(\pm)}}

\def\0 {\nonumber}

\begin{document} 
\setcounter{page}{0}
\newcommand{\inv}[1]{{#1}^{-1}} %inverse 
\renewcommand{\theequation}{\thesection.\arabic{equation}} 
\newcommand{\be}{\begin{equation}} 
\newcommand{\ee}{\end{equation}} 
\newcommand{\bea}{\begin{eqnarray}} 
\newcommand{\eea}{\end{eqnarray}} 
\newcommand{\re}[1]{(\ref{#1})} 
\newcommand{\qv}{\quad ,} 
\newcommand{\qp}{\quad .} 

\thispagestyle{empty}
%\begin{titlepage} 
%\title{} 
\begin{flushright} \small
UUITP-04/06 \\ HIP-2006-17/TH \\ 
\end{flushright}
\smallskip
\begin{center} \LARGE
{\bf A brief review of supersymmetric non-linear sigma models and generalized complex geometry}
\footnote{Contribution to ``The 26th Winter School GEOMETRY AND PHYSICS'', Czech Republic, Srni, January 14 - 21, 2006}¥
 \\[12mm] \normalsize
{\bf Ulf~Lindstr\"om$^{a,b}$,} \\[8mm]
 {\small\it
$^a$Department of Theoretical Physics 
Uppsala University, \\ Box 803, SE-751 08 Uppsala, Sweden \\
~\\
$^b$HIP-Helsinki Institute of Physics, University of Helsinki,\\
P.O. Box 64 FIN-00014  Suomi-Finland\\
~\\
}
\end{center}
\vspace{10mm}
\centerline{\bfseries Abstract} \bigskip

\noindent  
This is a review of the relation between supersymmetric non-linear sigma models and target space geometry.
In particular, we report on the derivation of  generalized K\"ahler geometry from sigma models with
additional spinorial superfields. Some of the results reviewed are: Generalized complex geometry from sigma models 
in the Lagrangian formulation; Coordinatization of generalized K\"ahler geometry in terms of chiral, twisted chiral 
and semi-chiral superfields; Generalized K\"ahler geometry from sigma models in the Hamiltonian formulation.
\eject
\normalsize

%\addtocontents{toc}
%\tableofcontents

%\end{titlepage}
\tableofcontents
\section{Introduction}
The construction of Generalized Complex Geometry (GCG), \cite{hitchinCY}, \cite{gualtieri} was 
motivated partly by understanding generalized Calabi-Yau manifolds relevant to string-physics. Since its 
introduction in mathematics, it has found
several additional applications in physics \cite{Alekseev:2004np}-\cite{Pestun:2006rj}.  Here we shall mainly be concerned with the special case of 
Generalized K\"ahler Geometry (GKG) \cite{gualtieri} and its relation to supersymmetric nonlinear sigma models.

The history of this line of investigation goes back 
to the original realization of the close relation between supersymmetric nonlinear sigma models and complex geometry,
some twenty-five years ago \cite{Zumino:1979et}, \cite{Alvarez-Gaume:1981hm}. The renewed interest is due to Gualtierei's proof that 
the bi-hermitean target-space geometry of Gates, Hull and Ro\v cek \cite{Gates:1984nk}
\footnote{See also \cite{Howe:1985pm}} may be mapped to GKG.  The mapping is nontrivial, however. 
In particular the sigma models are defined in terms of super-fields transforming in the tangent space $T$ 
of the target space $\kal{T}$, whereas the definition of GKG also requires the cotangent space $T^*$.
A direct relation between the sigma model and GKG thus requires additional $T^*$-fields.
Such a model is also of interest from another point of view; The second supersymmetry of \cite{Gates:1984nk}
only closes on-shell, in general. To obtain off-shell closure additional auxiliary superfields are needed.

This sets the scene for the investigation: The problem is to formulate a generalized $(1,1)$ nonlinear sigma 
model with superfields transforming in $T\oplus T^*$, require off-shell closure and read off the ensuing 
target space geometry. A related problem is that of interpreting the geometry for the manifest 
$(2,2)$ models. The present paper is a report on the results and difficulties of this program.

As the presentation is a review, we refer the reader to the references for details, but sections 2-5 present the background 
needed to appreciate the problem. Section 6 review results from the $N=(1,1)$ sigma model point of view, while section 7
takes the starting point in $N=(2,2)$ models, reporting on the recent result regarding coordinatization of, and the existence of a
 potential for generalized K\"ahler geometry. Finally, section 8 contains a brief summary of the recent understanding of generalized 
 K\"ahler geometry as the target space geometry in the Hamiltonian formulation of the sigma model.

\section{Sigma models}

A  $N=(p,q)$ two-dimensional supersymmetric nonlinear sigma model is a theory of maps $\phi^\mu$ from a 
supermanifold $\kal{M}^{(2|p,q)}$ to 
a target-space $\kal{T}$,
\beq
\phi^\mu: \kal{M}^{(2|p,q)}\longmapsto\kal{T}~,
\eeq{1}
found by minimizing the action
\beq
S=\int_{\kal{M}}\kal{L}(\phi)+\int_{\d\kal{M}}\kal{L}_{B}~,
\eeq{2}
where the form of the bulk Lagrange density $\kal{L}$ and the boundary term $\kal{L}_{B} $ depend 
on the number of supersymmetries. The boundary term is necessary is needed to preserve 
symmetries of the bulk-action in the open case. \footnote{In fact, there is an intricate interplay between the preservation 
of symmetries and the possible geometries of the sub-manifolds where the boundaries may lie (the $D$-branes) \cite{Albertsson:2001dv}-\cite{Howe:2005je}.}
We shall be concerned with closed sigma models in 
the two cases $N=(1,1)$ and  $\phi$ real, with\footnote{We use bosonic light-cone coordinates $\xi^\+$ and $\xi^=$.
 The double plus/minus
notation is in keeping with the $2d$ notation where a spinor $\theta$ has components $\theta^+$ and $\theta^-$.}
\beq
\kal{L}(\phi)=\d_{\+}\phi^\mu E_{\mu\nu}(\phi)\d_{=}\phi^\nu~,
\eeq{3}
and $N=(2,2)$ and $\phi$ complex, with 
\beq
\kal{L}(\phi)=K(\phi, \bar\phi)~.
\eeq{4}
In \enr{3}, the target space geometry enters these expressions through the target space metric $G_{\mu\nu}$ and the 
antisymmetric $B$-field  $B_{\mu\nu}$ in the combination\footnote{This is a slight abuse of notation, since the 
metric and  $B$-field are the lowest components of these superfield functions.}
\beq
E_{\mu\nu}(\phi)\equiv G_{\mu\nu}(\phi)+B_{\mu\nu}(\phi)~.
\eeq{5}
This also covers the $N=(2,2)$ action, since reducing \enr{4} to $N=(1,1)$ components yields 
special cases of \enr{3}, \enr{5} where the metric and  $B$-field are given as derivatives of the potential
$K(\phi, \bar\phi)$.  We thus see that more super-symmetry implies more restrictions on the target-space geometry.
In two dimensions the situation is (partly) summarized in the following table

\vskip .1in
\begin{table}[htbp]
\centering
\begin{tabular}{|l|c|c|c|c|c|}
\hline
Supersymmetry & (0,0) or (1,1) & (2,2) & (2,2) & (4,4) & (4,4) \\ 
\hline
Background & $G,B$ &  $G$ & $G,B$ & $G$ & $G,B$ \\
\hline
Geometry   &  Riemannian & K\"ahler & bihermitian & hyperk\"ahler & bihypercomplex \\
\hline
\end{tabular}
\caption{The geometries of sigma-models with different supersymmetries.}
\end{table}
\vskip -.1in

We shall focus on the $(2,2)$ case which we now briefly describe from an $N=(1,1)$ superspace point of view.

\section{The Gates-Hull-Ro\v cek geometry}
\label{GHR}

Starting from the action \enr{2} with the Lagrange density \enr{3}, we may ask under which conditions
there is an additional left and an additional right supersymmetry
\beq
\delta \phi^\mu=\e^+J^{\mu}_{(+)\nu}D_{+}\phi^\nu+\e^-J^{\mu}_{(-)\nu}D_{-}\phi^\nu~,
\eeq{6}
where $D_{\pm}$ are the $N=(1,1)$ superspace covariant spinorial derivatives. In \cite{Gates:1984nk} the answer 
to this was found to be that the target space geometry has to satisfy the following requirements;\\
(i) The tensors $J_{(\pm)}(\phi)$ have to be complex structures, i.e.,  $J^2_{(\pm)}=-1$ and $ \kal{N}_{(\pm)}=0$, where 
$\kal{N}_{(\pm)}$ denotes the corresponding Nijenhuis tensors.\\
(ii) The tensors $J_{(\pm)}(\phi)$ also have to be covariantly constant with respect to  torsionful connections,
$\nabla^{(\pm)}J_{(\pm)}=0$ where $\G^{(\pm)}\equiv \G^{(0)}\pm G^{-1}H$, $\G^{(0)}$ is the Levi-Civita connection
 and $H=dB$.\\
(iii) The metric G is hermitean with respect to both complex structures, $J_{(\pm)}^tGJ_{(\pm)}=G$.\\
(iv) The complex structures preserve the torsion, i.e.,  $J_{(\pm)}HJ_{(\pm)}=J_{(\pm)}H$.\\

The conditions (i)-(iv) are a consequence of invariance of the action \enr{2} under the second supersymmetry \enr{6} and of 
closure of the algebra of that symmetry. It is an important fact that the algebra only closes on-shell in general. If the two complex 
structures  commute it does close off-shell, however, and there is a manifest $(2,2 )$ description of the sigma model in terms of chiral and 
twisted chiral  $(2,2 )$ superfields. 

In \cite{Lyakhovich:2002kc}, the conditions (i)-(iv) were reformulated  in terms of the Poisson structures 
\beq
\pi_{(\pm)}\equiv (J_{(+)}\pm J_{(-)} )G^{-1}~,
\eeq{7}
a reformulation which anticipates the subsequent  generalized complex geometry description.  Before we turn to
the equivalence between the bihermitean geometry and the GCG, we need to briefly introduce the latter.

\section{Generalized complex geometry}
\label{GCG}

Generalized complex geometry is introduced in \cite{hitchinCY} and elaborated on in \cite{gualtieri}. A coordinate 
formulation useful for physicists is given in \cite{Lindstrom:2004iw}. Here we only recapitulate a few important facts.

The basic object in the definition of a GCG is the generalized complex structure $\kal{J}$. This is introduced in a manner which mimics 
the description of  a complex structure, namely as an automorphism of the sum of the tangent space and the cotangent space
\beq
\kal{J}: T\oplus T^*\mapsto  T\oplus T^*
\eeq{8}
which squares to minus one
\beq
\kal{J}^2=-1~.
\eeq{9}
The projection operators
\beq
\Pi_{\pm}\equiv\half(1\pm i\kal{J})
\eeq{10}
are used to define integrability by requiring that
\beq
\Pi_{\mp}[\Pi_{\pm}(X+\xi),\Pi_{\pm}(Y+\eta)]_{c}=0~,
\eeq{11}
where $X+\xi,Y+\eta \in T\oplus T^*$ and the bracket is the Courant bracket defined by\footnote{In the presence of a closed three-form $H$,
the Courant bracket can be 
modified (twisted) by adding a term $\imath_{X}\imath_{Y}H$}
\beq
[X+\xi , Y+\eta]_{c}\equiv [X,Y]+\pounds_{X}\eta-\pounds_{Y}\xi
-\frac 1 2  d(\imath _{X}\eta -\imath _{Y}\xi)~.
\eeq{12}
For physics it is highly relevant that the group of automorphisms of the Courant bracket apart from diffeomorphisms
also includes b-ransforms, defined by
\beq
 e^b(X+\xi)\equiv X+\xi+\imath _{X}b~,
 \eeq{16}
 for a closed two-form $b$.
 
As a final ingredient, the natural pairing $\kal{I}$, defined by
\beq
<X+\xi, Y+\eta >=\imath _{X}\eta +\imath _{Y}\xi~,
\eeq{13}
is required to be hermitean with respect to $\kal{J}$;
\beq
\kal{J}^t\kal{I}\kal{J}=\kal{I}~.
\eeq{14}
Note that, apart from the last condition, the definition of an ordinary complex structure is recovered 
by replacing the sum of the tangent an cotangent spaces by the tangent space only and the Courant bracket by the 
Lie bracket.

In the local basis $(\partial_{\mu} ,dx^\nu)$, the generalized complex structure and the pairing metric take the form
\beq
{\cal{J}}=\left(\begin{array}{cc}
J&P\cr
L&K\end{array}\right)~,\quad {\cal{I}}=\left(\begin{array}{cc}
0&1_{d}\cr
1_{d}&0\end{array}\right)~,
\eeq{15}
where $d$ is the dimension of the manifold and the blocks are maps between the possible combinations of $T$ and $T^*$.
A $b$-transform acts on $\kal{J}$ according to
\beq
{\cal{J}}_{b}=\left(\begin{array}{cc}
1&0\cr
b&1\end{array}\right){\cal{J}}\left(\begin{array}{cc}
1&0\cr
-b&1\end{array}\right).
\eeq{17}
To appreciate the scope of GCG, it is useful to note that a $b$-transform may take us from a $\kal{J}$ representing ordinary 
complex geometry to one representing symplectic geometry. Generalized complex geometry contains both types of geometries 
as special cases.

An important special case of a GCG, called a generalized K\"ahler geometry (GKG), is defined in \cite{gualtieri}. It 
involves two commuting generalized complex structures $\kal{J}_{1}$ and $\kal{J}_{2}$ and requires that the metric formed 
from these,
\beq
\kal{G}\equiv -\kal{J}_{1}\kal{J}_{2}~,
\eeq{18}
is positive definite. When $\kal{J}_{1}$ represents the ordinary complex structure of K\"ahler geometry and  $\kal{J}_{2}$ 
represents the K\"ahler form, then $\kal{G}$ is made from the K\"ahler metric.
It is this generalized K\"ahler geometry which is of interest for the sigma models, and we now turn to that relation.

\section{Sigma models realizations}

In \cite{gualtieri} it is shown that there is a map between the bihermitean geometry of section \ref{GHR} and the 
generalized complex geometry of section \ref{GCG}.
The geometric data $J_{(\pm)}, G, B$ always defines a GKG with generalized complex structures
\beq
 {\cal J}_{1,2} = - \half
 \left ( \begin{array}{ll}
       J_{(+)} \pm J_{(-)}& -(\omega_{(+)}^{-1} \mp \omega_{(-)}^{-1}) \\
     \omega_{(+)} \mp \omega_{(-)} & - (J_{(+)}^t \pm J^t_{(-)})~,
\end{array} \right )
\eeq{19}
where $\omega_{(\pm)}$ are the symplectic forms corresponding to $J_{(\pm)}$.
Up to $b$-transforms and diffeomorphims, the inverse is also true. We emphasize that the relation is independent of an actual sigma model realization, closure of 
the algebra et c. This relation, however, tells us that if we add fields transforming as co-vectors of the target space $\kal{T}$ we 
have a chance of realizing the GKG directly from a (generalized) sigma model. To keep the physical degrees of freedom the same as in 
the original sigma model, the new fields should be auxiliary. This is also what we expect if the algebra of non-manifest supersymmetries 
is to close in the generalized model.

In fact, if we turn to the $(2.2)$ models and their reduction to $(1,1)$ we discover some of the necessary structure. 
Namely, a $(2.2)$ sigma model written in terms of semi(anti) chiral superfields 
\footnote{Introduced in \cite{Buscher:1987uw}.} $\mathbb{X} ,\overline{\mathbb{X}}$ will contain spinorial auxiliary fields when reduced to $(1,1)$. This is
an example of the kind of models we are looking for, and will be discussed in more detail below. A $(1,1)$ generalized sigma model with
auxiliary spinorial superfields $ S_{\pm \mu}$ transforming in $T^*$ reads \cite{Lindstrom:2004eh}
\beq
S=\int_{\kal{M}}d^2\xi d^2\theta\left(S_{+\mu}E^{\mu\nu}(\phi)S_{-\nu}+S_{(+\mu}D_{-)}\phi^\mu\right)~.
\eeq{20}
Here $E^{\mu\nu}$ is the inverse of $E_{\mu\nu}$ defined in \enr{5}, but this condition may be relaxed and $S$ may be studied
on its owen with no assumption of invertibility.
The most general ansatz for the second supersymmetry  (in terms of left and right transformations) reads
\ber
\delta ^{(\pm)}\phi^\mu&=&\epsilon^{\pm}\left({ D_{\pm}\phi^\nu J^{(\pm)\mu}_{~~\nu}}-S_{\pm\nu}P^{(\pm)\mu\nu}\right)\cr
\delta^{(\pm)} S_{\pm\mu}&=&\epsilon^{\pm}\left(i\partial_{\pp}\phi^\nu L^{(\pm)}_{\mu\nu}-D_{\pm}S_{\pm\nu}K^{(\pm) \nu}_{\mu}
+S_{\pm\nu}S_{\pm\sigma}N^{(\pm)\nu\sigma}_{\mu}\right.\cr
&&\left. \qquad \qquad +D_{\pm}\phi^\nu D_{\pm}\phi^\rho M^{(\pm)}_{\mu\nu\rho}+D_{\pm}\phi^\nu S_{\pm\sigma}
Q^{(\pm)\sigma}_{\mu\nu}\right)\cr
\delta^{(\pm)} S_{\mp\mu}&=&\epsilon^{\pm}\left(D_{\pm}S_{\mp\nu}R^{(\pm)\nu}_{\mu}+
D_{\mp}S_{\pm\nu}Z^{(\pm)\nu}_{\mu}+D_{\pm}D_{\mp}\phi^\nu T^{(\pm)}_{\mu\nu}\right.\cr
&&\left. \qquad \qquad +S_{\pm\rho}D_{\mp}\phi^\nu U^{(\pm)\rho}_{\mu\nu}
+D_{\pm}\phi^\nu S_{\mp\rho} V^{(\pm)\rho}_{\mu\nu}\right.\cr
&&\left. \qquad \qquad +D_{\pm}\phi^\nu D_{\mp}\phi^\rho X^{(\pm)}_{\mu\nu\rho}
+S_{\pm\nu}S_{\mp\rho}Y^{(\pm)\nu\rho}_{\mu}\right)~,
\eer{21}
a considerable complication as compared to the case with no auxiliary fields  \enr{6} above which contains only the first 
term on the right hand side of the first row. The many higher rank tensors and the many differential and algebraic equations resulting from invariance of
the action \enr{20} and closure of the algebra, make the general analysis cumbersome.\footnote{In \cite{Lindstrom:2004eh} a solution for the transformations
\enr{21} respecting certain discrete symmetries of \enr{20} was found . This solution was extended to the $G=0$ case in \cite{Bergamin:2004sk}.}
We next discuss some examples and some simplifications.

\section {Examples}

In this section we discuss sigma model realization of generalized complex geometry in some examples. 
It constitutes a brief recapitulation of the results of \cite{Lindstrom:2004iw} and \cite{Bredthauer:2006hf}.

\subsection{$(1,0) \to (2,0)$: A toy model.}

In \cite{Lindstrom:2004iw} we study the simpler question of just one extra (left) supersymmetry, going from
$N=(1,0)$ to $N=(2,0)$. The first case we look at is the toy model defined by the action
\beq
S=\int_{\kal{M}}d^2\xi d\theta S_{+\mu}\partial_{=}\phi^\mu~.
\eeq{22}
This is a topological model, whose second supersymmetry reads
\ber
&&\delta\phi^\mu=\epsilon^+(D_{+}\phi^\nu J^\mu_{~\nu}-S_{+\nu}P^{\mu\nu})\cr
&&\delta S_{+\mu}=\epsilon^+\left(\partial_{\+}\phi^\nu L_{\mu\nu}-D_{+}S_{+\nu}K^\nu_{\mu}
+S_{+\nu}S_{+\rho}N^{~\nu\rho}_{\mu}\right.\cr
&&\left.+D_{+}\phi^{\nu}D_{+}\phi^{\rho}M_{\mu\nu\rho}
+D_{+}\phi^{\rho}S_{+\nu}Q_{\mu\rho}^{~~\nu}\right)
\eer{23}
We show that invariance of the action \enr{22} under the transformations \enr{23} and closure of the algebra of those transformations
is equivalent to
\beq
{\cal{J}}=\left(\begin{array}{cc}
J&P\cr
L&K\end{array}\right)~,
\eeq{24}
being a generalized complex structure. The higher rank tensors in \enr{23} are expressible as derivatives of those occurring in \enr{24}.
Hence, in this case the model exactly determines the target space geometry to be GCG. In fact  this still holds true if we add a Wess-Zumino 
term, provided that the Courant bracket is modified to its twisted version.

\subsection{$(1,0) \to (2,0)$: The sigma model.}

The $(1,0)$ sigma model action reads
\beq
S=\int_{\kal{M}}d^2\xi d\theta(D_{+}\phi^\mu S_{=\mu}
-S_{+\mu}\partial_{=}\phi^\mu-S_{+\mu}S_{=\nu}E^{\mu\nu})Ô,
\eeq{25}
where the auxiliary fields are one spinor and one vector field.
The ansatz for the second supersymmetry is given by expressions which is are obvious modifications
of \enr{21}, in particular they define the tensors $J,P,L,K,R,T$ and $Z$.

Here the analysis of the algebraic and differential conditions is already very involved. In \cite{Lindstrom:2004iw}
we are able to show that they are satisfied for generalized complex geometry, and that they precisely correspond to 
GCG under some additional weak assumptions, but we fail to show that they lead uniquely to GCG.

Perhaps this is not too surprising, given the many  first order actions possible. E.g., for the $(1,1)$ model, the action 
\enr{20} may be replaced by
\beq
S = \int d^2\xi d^2\theta 
(\tilde S_{+\mu}\tilde E^{\mu\nu}\tilde S_{-\nu}+D_+\phi^\mu E_{\mu\nu} D_-\phi^\nu) ~,
\bigskip
\eeq{26}
and when $\tilde E^{\mu\nu}$ is invertible the two are equivalent (related via field redefinitions). There is also the 
question of canonical coordinates, i.e., coordinates where the natural pairing metric takes the form \enr{15}.
The action \enr{22} turned out to be formulated in such coordinates, but in principle the fields in \enr{25} might be 
related to the canonical coordinates via a coordinate transformation in $T\kal{T}\oplus T^*\kal{T}$.

One additional interesting observation made in \cite{Lindstrom:2004iw} is that the algebraic conditions for the model
\enr{25} may be summarized in a formula for a $3d\times 3d$ matrix;
\beq
\left(\begin{array}{ccc}
R&T&Z\cr
0&{J}&{P}\cr
0&{L}&{K}\end{array}\right)^2=-1_{3d}~.
\eeq{27}
This indicates that the underlying geometry of the general model \enr{20} might be best formulated using {\em two} copies of 
the cotangent space, rather than one. This hypothesis is tested in the next example.

\subsection{$(1,1) \to (2,2)$ The symplectic model}

In \cite{Bredthauer:2005zx} we study a special case of the action \enr{26}, namely the case when the target space 
metric $G_{\mu\nu}$ is zero, i.e.,
\beq
S = \int d^2\xi d^2\theta 
(S_{+\mu}\Pi^{\mu\nu}S_{-\nu}+D_+\phi^\mu B_{\mu\nu} D_-\phi^\nu)~,
\eeq{28}
with $\Pi^{\mu\nu}$ antisymmetric and invertible. (Without the assumption of invertibility, this 
is a $(1,1)$ Poisson sigma model.) We find that the full target space geometry has a neat formulation in terms of 
$3d\times 3d$ matrices defined on $\kal{M}\equiv T{\cal{T}}\oplus(T^*{\cal{T}}_+\oplus T^*{\cal{T}}_-)$, where plus 
and minus just label two copies of the cotangent bundle. 
The condition \enr{27} now translates into the existence of the two commuting (almost) complex structures
\beq
  {\bf J}^{\p} = \left(\begin{array}{ccc}
             \ph{-}J^\p&-P^\p&\ph{-}0 \cr
             -L^\p&\ph{-}K^\p&\ph{-}0 \cr
             \ph{-}T^\p&-Z^\p&\ph{-}R^\p
             \end{array}\right)
             \quad
  {\bf J}^\m = \left(\begin{array}{ccc}
             \ph{-}J^\m&\ph{-}0&-P^\m \cr
             \ph{-}T^\m&\ph{-}R^\m&-Z^\m \cr
             -L^\m&\ph{-}0&\ph{-}K^\m
            \end{array}\right)~,
\eeq{30}
with respect to which the (degenerate) ``metric''
\ber
{\bf G} = {\bf G}^t =
             \half  \left(\begin{array}{ccc}
\ph{-}0&\ph{-}0&\ph{-}0 \cr
             \ph{-}0&\ph{-}0&\ph{-}\Pi \cr
             \ph{-}0&\ph{-}\Pi^t&\ph{-}0\end{array}\right)~,
\eer{29}
is hermitean
\beq
{\bf J}^{\ppm t}{\bf G} {\bf J}^{\ppm} = {\bf G}~,\quad
  {\bf J}^{\ppm2} = - {\bf 1}~, \quad
  {[}{\bf J}^\p,{\bf J}^\m] = 0~.
  \eeq{31}
This looks a lot like the bi-hermitean geometry of section \ref{GHR}, only lifted to $\kal{M}$.
Indeed, we also define a (torsion-free, flat and diagonal) connection $\bf\G$
such that 
\beq
\bf\nabla{\bf J}^{\ppm} = 0 ~.
\eeq{32}
In fact, taken into account the degeneracy due to $G_{\mu\nu}=0$, we do recover a lifted version of the bi-hermitean geometry.
The details may be found in  \cite{Bredthauer:2005zx}, here we just note that the $b$-transform also arises  naturally
in this context; it is part of the gauge-transformations related to the connection $\bf\G$.
It should be interesting to extend this analysis to no-zero $G_{\mu\nu}$ and, if possible, to elucidate the relation to GCG.

\section{The manifest $N=(2,2)$ models}

The discussion so far has been concerned with finding the restrictions on the target-space that result from requiring 
invariance of an $N=(1,1)$ action under additional supersymmetries as well as closure of the corresponding algebra.
When the algebra closes it should in principle be possible to find a $(2,2)$ superspace formulation. This leads to the 
question of what  is the most general $(2,2)$ superfield formulation that corresponds to the $(1,1)$ sigma models.
As may be inferred from the preceding discussion of these models, this is tantamount to asking how to coordinatize 
$ker[J_{(+)},J_{(-)}]\oplus coker[J_{(+)},J_{(-)}]$ using $(2,2)$ superfields. (The reason for this split is that we 
already know how to coordinatize the kernel, as mentioned in section \ref{GHR}.)  The mathematical problem of 
how to choose coordinates for this space has been around for a long time  \cite{Sevrin:1996jr} 
\cite{Grisaru:1997pg}, but was resolved only recently in \cite{Lindstrom:2005zr}. Before stating the results of that 
paper we have to define the relevant $(2,2)$ superfields.

\subsection{$N=(2,2)$ superfields}
\label{fields}

Denoting the $2d$ spinorial $(2,2)$ covariant derivatives by $\mathbb{D}_{\pm}$ and 
$\overline{\mathbb{D}}_{\pm}$, the following constraints characterize the $(2,2)$ superfields:\\
Chiral $\phi$, anti-chiral $\bar \phi$, 
$(2,2)$ superfields are defined by 
\beq
\overline{\mathbb{D}}_{\pm}\phi=\mathbb{D}_{\pm}\bar\phi=0~.
\eeq{33}
 Twisted chiral $\chi$ and twisted antichiral $\bar\chi$ superfields are defined by
 \beq
\mathbb{D}_{+}\chi=\overline{\mathbb{D}}_{-}\chi=\overline{\mathbb{D}}_{+}\bar\chi=
\mathbb{D}_{-}\bar \chi=0~.
\eeq{34}
Left or right semi-chiral  $\mathbb{X}_{L,R}$ and left or right anti semi-chiral $\overline{\mathbb{X}}_{L,R}$ 
superfields are defined by \cite{Buscher:1987uw}:
\beq
\bar {\mathbb{D}}_{+}\mathbb{X}_L=0\quad \mathbb{D}_{-}\bar{\mathbb{X}}_R=0~,
\eeq{35}
and the hermitean conjugate relations. 

\subsection{$N=(2,2)$ actions and their reduction}

A general $N=(2,2)$ action involving the fields defined in subsection \ref{fields} reads

\beq
S=\int d^2\xi d^2\theta d^2\bar\theta 
K (\phi,\bar\phi,\chi,\bar\chi,\mathbb{X}_L,\overline{\mathbb{X}}_L,\mathbb{X}_R,\overline{\mathbb{X}}_R)~.
\eeq{36}

To compare this action to the $(1,1)$ action \enr{20}, we need to reduce it to $(1,1)$ superfield form. To this end
we define the $(1,1)$ covariant derivatives $D_{\pm}$ and second supersymmetry charges
\beq
D_{\pm}=\mathbb{D}_{\pm}+\bar {\mathbb{D}}_{\pm}~,
\quad Q_{\pm}=i(\mathbb{D}_{\pm}-\bar {\mathbb{D}}_{\pm})~,
\eeq{37}
and the $(1,1)$ component fields
\ber
&\phi\equiv (\phi_{1}+i\phi_{2})|&X_{L}\equiv \mathbb{X}_L|\qquad \Psi_{L-}\equiv Q_{-}\mathbb{X}_L |\cr
&\chi\equiv (\chi_{1}+i\chi_{2})|&X_{R}\equiv \mathbb{X}_R|\qquad \Psi _{R+}\equiv Q_{+}\mathbb{X}_R | ~,
\eer{38}
where the vertical bar denotes setting the second fermi-coordinate, $\th^2$, to zero. The reduction is then achieved by writing
the Lagrangian as
\beq
D^2Q^2K (\phi,\bar\phi,\chi,\bar\chi,\mathbb{X}_L,\overline{\mathbb{X}}_L,\mathbb{X}_R,\overline{\mathbb{X}}_R)|
=D^2K(\phi_{i},\chi_{i}, X_{L,R}, \Psi_{L-},\Psi_{R+})~.
\eeq{39}
The set $\phi_{i},\chi_{i}, X_{L,R}$, $i=1,2$ is identified with the scalar fields $\phi$ in \enr{20}, while
the auxiliary fields $\Psi_{L-}, \Psi_{R+}$ still need a redefinition\footnote{As can be seen from the number of auxiliary spinors, this is not 
the general case of \enr{20}, however.}. The details of this may be found in \cite{Lindstrom:2004hi}. Eliminating the auxiliary fields then bring us to 
the class of sigma models studied in \cite{Gates:1984nk}  before we have auxiliary spinors transforming in
$T^*$.

In \cite{Gates:1984nk} and in \cite{Ivanov:1994ec}, it is shown that precisely when the two complex 
structures commute, $[J_{(+)},J_{(-)}]=0$, there is a $(2,2)$ description of the sigma model in terms of chiral and twisted 
chiral superfields. I.e., $ker(J_{(+)}+J_{(-)})\oplus ker (J_{(+)}-J_{(-)})$ is precisely coordinatized by those fields.
(The split of the kernel of the commutator corresponds to the two types of fields.)  This is the case when all semi-chiral fields 
are set to zero in \enr{36}:
\beq
 S\to\int d^2\xi d^2\theta d^2\bar\theta K(\phi,\bar\phi,\chi,\bar\chi)~.
\eeq{40}
The question of the co-kernel of the commutator now arises. Is it completely described by turning on the semi chiral fields
or are there still other $(2,2)$ fields needed?  The answer, given in \cite{Lindstrom:2005zr} is that this is indeed enough.
It follows that {\em the full generalized K\"ahler geometry may be described by coordinates that are chiral, twisted chiral and 
semi-chiral superfields} and that {\em the GKG has a potential $(K)$ which determines the metric and the $B$-field }
(in a non-linear manner). A surprising result is further that $K$ has an interpretation as a generating function for 
certain symplectomorphisms.  Important ingredients in this derivation are the reformulation in \cite{Lyakhovich:2002kc}
of the bi-hermitean constraints in 
terms of the Poisson structures \enr{7} and the introduction in \cite{hitchinP} of a third Poisson structure 
\beq
\sigma :=[J_{+},J_{-}]G^{-1}~.
\eeq{41}

\section{Recent development}

In looking for the target space geometry of the generalized sigma model \enr{20}, or \enr{26} we face the problem of
non-uniqueness of the auxiliary field coupling. In fact, starting from the $(2,2)$ form \enr{36} we find only a sub-class of 
the actions described by \enr{20}. The fact that the auxiliary fields are world-sheet spinors also doubles the number of degrees
of freedom that we need to describe the cotangent bundle. So far these problems have prevented a complete determination
of the target space for the action \enr{20}, and there are even hints that the geometry it corresponds to may be larger than
GCG. It is thus gratifying that there exists a Hamiltonian approach to the sigma models where these problems are largely overcome.
Briefly, the Hamiltonian for the sigma model will contain only one ``extra«« field rather than two and  the form of the Hamiltonian is
essentially fixed.

This development was initiated in \cite{Zabzine:2005qf}, where it is shown that there is a direct relation between generalized complex 
geometry and a the Hamiltonian formulation of sigma models. This discussion is model independent, and the particular case of
certain poisson sigma models has since been  discussed in \cite{Calvo:2005ww}. 

Recently this development was completed in \cite{Bredthauer:2006hf} where we show that generalized K\"ahler geometry is  precisely
the target space geometry when you require a second closing supersymmetry in the Hamiltonian formulation. From this formulation we also derive the  
correspondence \enr{19} to the bi-hermitean geometry of \cite{Gates:1984nk}.

\vspace{2cm}

\noindent{\bf Acknowledgement}: I want to express my gratitude to all my collaborators on the papers discussed, Andreas Bredhauer, 
Jonas Persson, Ruben Minasian, Martin Ro\v cek, Alessandro Tomasiello, Rikard von Unge and in particular
to Maxim Zabzine for close collaboration on the subject of this paper over several years.
The research is supported by VR grant 621-2003-3454  Partial support for this research is further provided  by EU grant (Superstring theory)
MRTN-2004-512194.

\eject

\end{document}